\documentclass[10pt, superscriptaddress, 
    prl, aps, twocolumn]{revtex4-2}

\usepackage{orcidlink, fontawesome, amssymb}

\newcommand{\IoA}{
    Institute of Astronomy, 
    University of Cambridge, 
    Madingley Road, 
    Cambridge, CB3 0HA, UK}
\newcommand{\KICC}{
    Kavli Institute for Cosmology, 
    University of Cambridge, 
    Madingley Road, 
    Cambridge, CB3 0HA, UK}
\newcommand{\DAMTP}{
    Department of Applied Mathematics 
    and Theoretical Physics, 
    Centre for Mathematical Sciences, 
    University of Cambridge, 
    Wilberforce Road, 
    Cambridge, CB3 0WA, UK}

\begin{document}

\title{
    On the Interpretation of Tests of the Horizon Area Law
    }

\author{Selma Vangstein\, 
    \orcidlink{0009-0006-8612-0995}}
\email{snv28@cam.ac.uk}
\affiliation{\IoA}

\author{Christopher J.\ Moore\, 
    \orcidlink{0000-0002-2527-0213}}
\affiliation{\IoA} 
\affiliation{\DAMTP} 
\affiliation{\KICC}

\date{\today}

\begin{abstract}
    This Letter investigates gravitational-wave tests of the horizon area law.
    Although they provide a striking consistency check of general relativity (GR), these tests rely on models that assume GR, making their interpretation inherently model dependent.
    We show with a specific example---simulations of merging boson stars---that exotic compact objects can mimic merging black holes and pass an area-law test at a confidence as high as ${\sim \! 92\%}$ despite not even having event horizons.
\end{abstract}

\maketitle

{\textit{\textbf{Introduction}}---}Gravitational waves (GWs) enable tests of general relativity (GR) with strong and dynamical gravitational fields \cite{TGR_review_1, TGR_review_2}.
Such tests began with the first GW observation \cite{GW150914, TGR_GW150914} and have since been applied to larger catalogs \cite{2026arXiv260719293T}.
Most tests use signals from merging black holes (BHs),
a notable exception being the stringent constraint on the speed of GWs from the joint observation of the neutron star merger GW170817 and the associated gamma-ray burst 170817A \cite{GW170817, TGR_GW170817}.

A recent high-profile example is a test of the horizon area law using GW250114\_082203 (hereafter GW250114), the loudest GW signal to date \cite{GW250114_area_test}.
Previous area-law tests using the quieter events GW150914 \cite{GW150914_area_test} and GW230814\_230901 \cite{GW230814_area_test} did not reach the same significance.
In classical GR, Hawking's area theorem \cite{Hawking} states that (subject to certain standard assumptions) the total area of the event horizons in a system of BHs is non-decreasing with time and increases during major events such as mergers.
The theorem plays a central role in BH thermodynamics \cite{Walkd_BH_thermo}, the second law \cite{Bardeen_four_laws, Bekenstein_second_law}, Hawking radiation \cite{Hawking_BH_explosions}, and the information loss problem \cite{Almheiri_2013_info}.

Reference \cite{GW250114_area_test} tested the area law by splitting the GW250114 signal into inspiral and ringdown portions, with a gap around merger. 
The inspiral was analyzed using a binary BH model parameterized by the component masses and spins.
The ringdown was analyzed separately using a sum of damped sinusoids whose frequencies and damping times were fixed to the Kerr quasinormal-mode (QNM) values determined by the remnant mass and spin.
Both models assume Kerr BHs in vacuum GR \cite{Kerr_1963}; we refer to this restrictive set of assumptions as \emph{Kerr-vacuum-GR}.
The horizon areas of all three objects were calculated using the Kerr horizon-area formula \cite{Bardeen_four_laws}:
\begin{equation} \label{eq:Kerr_horizon_area}
    A = 8\pi M^2 (1+\sqrt{1-|\vec{\chi}|^2}) ,
\end{equation}
where $M$ and $\vec{\chi}$ are the BH mass and dimensionless spin.
Comparing the initial area inferred from the inspiral with the final area inferred from the ringdown yielded strong evidence for an increase during the merger.

This constitutes an impressive internal consistency check of GR.
However, it is unclear if a specific aspect of the theory is being tested.
Horizon areas are not measured directly. 
Instead, masses and spins are inferred using parameterized models that assume Kerr-vacuum-GR, and horizon areas are computed via Eq.~\ref{eq:Kerr_horizon_area}.

This raises a natural question: to what extent should we be surprised or impressed that analyses that assume GR yield results consistent with GR?
More pointedly, could systems that violate the Kerr-vacuum-GR assumptions, perhaps even systems with very substantial violations, still pass this consistency check?
We address these questions using a class of exotic compact objects---boson stars (BSs)~\cite{Kaup_1968_geon, Friedberg_1987_soliton_stars, Kleihaus_2005_Qballs}---as a case study.

BSs are hypothetical compact objects composed of scalar fields that lack event horizons.
Therefore, BSs violate the first two of the Kerr-vacuum-GR assumptions. 
They do, however, retain the third: their dynamics are governed by GR and therefore Hawking's area theorem continues to hold for simulation of BS mergers.
BSs exists with any mass and can have sufficiently high compactness, ${C \approx \mathrm{mass}/\mathrm{radius}}$, to effective mimic BHs \cite{Cardoso_Pani_2019, Bambi_2025_BH_mimickers}. 
Binary BS mergers provide a theoretically well-posed, non-Kerr-vacuum-GR scenario that can be studied with full numerical relativity (NR).
They are among the few BH mimickers for which such simulations exist.

In this Letter, the GW signal from an NR simulation of a binary BS is injected into simulated LIGO noise.
The binary parameters were chosen to match those of GW250114.
The signal is split into inspiral and ringdown portions and each is analyzed separately using Kerr-vacuum-GR models.
We closely follow the analysis in Ref.~\cite{GW250114_area_test}, combining the inspiral and ringdown results into a test of the area law.
Despite the injection strongly violating the underlying assumptions---indeed, no horizons are present before merger---the signal still passes the area-law test with a confidence as high as ${\sim \! 92\%}$.

We use geometrized units, ${G=c=1}$. 
For isolated BSs, the only length scales are the reduced Compton wavelength of the scalar field, ${\lambdabar=\hbar/\mu}$, and the BS mass $m$; only one can be specified independently (see below).

{\textit{\textbf{Boson Star Mergers}}---}BSs are composed of a minimally coupled complex scalar field $\varphi$. 
The action is
\begin{equation} \label{eq:action}
    S = \int \frac{\sqrt{-g}}{2} \bigg\{\frac{R}{8 \pi} - \big[\hbar^2 g^{\mu \nu} \nabla_{\mu} \bar{\varphi} \nabla_{\nu} \varphi + V(\varphi) \big] \bigg\} \mathrm{d}^4 x .
\end{equation}
The scalar field has mass $\mu$ and a solitonic potential \cite{Friedberg_1976_solitonic}
\begin{equation} \label{eq:potential}
    V(\varphi) = \mu^2|\varphi|^2 \left( 1 - 2|\varphi|^2/\sigma_0^{2} \right)^2 ,
\end{equation}
where ${\sigma_0=0.2}$.
Varying with respect to the metric and scalar field yields the Einstein-Klein-Gordon equations.

\begin{figure}[t]
    \centering
    \includegraphics[width=\columnwidth]{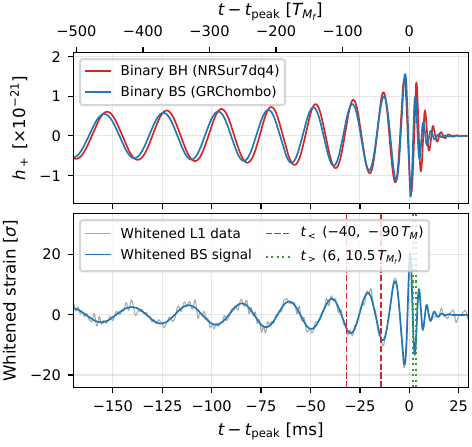}
    \caption{ \label{fig:injection} 
        The binary BS injection. 
        \emph{Top panel:} the plus polarization of the injected waveform shown alongside a binary-BH waveform with the same total mass, mass ratio $q=1$ and zero spins.
        \emph{Bottom panel:} the whitened data and injected signal in the L1 detector with the two inspiral end times $t_<$ and ringdown start times $t_>$ indicated. 
        Times are shown from $t_{\rm peak}$ in milliseconds and in units of the detector-frame final mass ${T_{M_f}}$ (see Tab.~2 of the supplement of Ref.~\cite{GW250114_area_test}).
    }
\end{figure}

Initial data for spherically symmetric, isolated BSs are obtained using the ansatz $\varphi(t, r) = A(r) e^{i(\delta \psi \pm \omega t)}$.
The central scalar amplitude, ${A=A(0)}$, is freely specified, but the frequency $\omega$ is determined by integrating outward using a shooting method \cite{Helfer_2022_shooting}.
We consider the relatively compact \texttt{A17} BS \cite{Evstafyeva_2024_BScatalog} with ${A=0.17}$.
Its compactness, ${C \approx 0.2}$, lies between that of neutron stars (${C \approx 0.1}$) and BHs (${C \approx 0.5}$).
This compactness is not particularly extreme, but, as we will show, it is sufficient for these BSs to act as reasonable BH mimickers when they merge.
We note that more compact BS solutions \cite{Collodel_2022_thin_shell, Marks_2025_ultracompact} are expected to be even better BH mimickers \cite{Cardoso_2016, Cardoso_2017, Cunha_2018}.

\begin{figure}[t]
    \centering
    \includegraphics[width=\columnwidth]{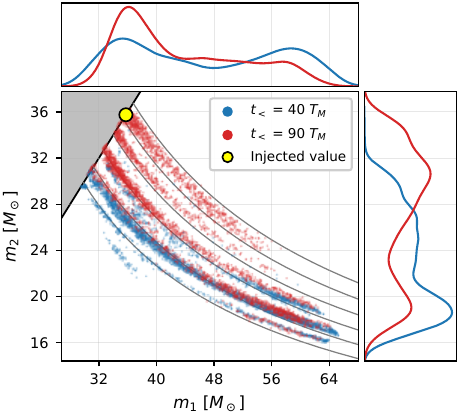}
    \caption{ \label{fig:inspiral} 
        The inspiral analysis.
        Posteriors on the component masses contain multiple ridges along contours of constant chirp mass (thin black lines).
        The two-dimensional posterior is represented using a cloud of sample points (as opposed to with a kernel density estimate) to avoid oversmoothing and obscuring the ridged posterior structure. 
        The injected BS masses are indicated by the yellow dot.
    }
\end{figure}

\begin{figure*}[t]
    \centering
    \includegraphics[width=0.96\textwidth]{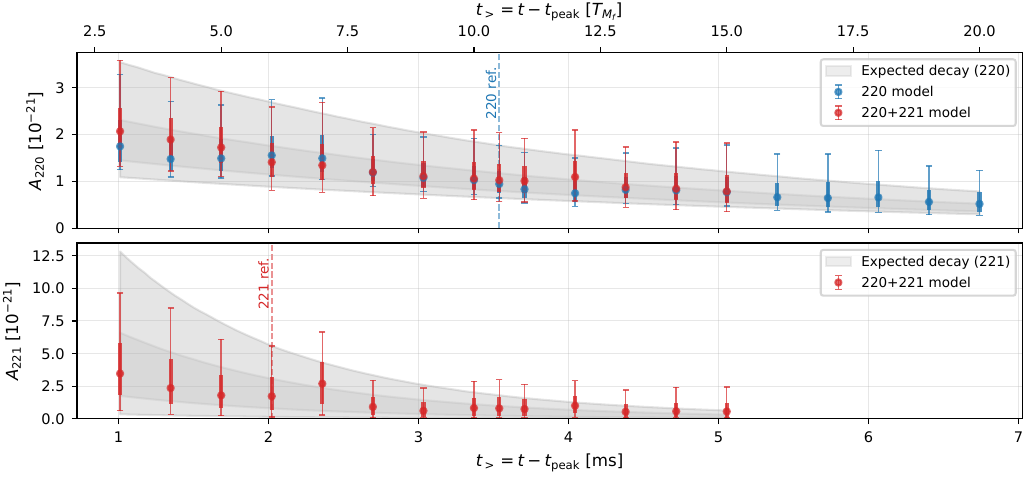}
    \caption{ \label{fig:ringdown} 
        The ringdown analysis. 
        Posteriors on the QNM amplitudes at different ringdown start times.
        This should be compared with Fig.~3 in Ref.~\cite{GW250114_area_test}.
        \emph{Top panel:} the amplitude of the fundamental ${(\ell,|m|,n)=(2,2,0)}$ QNM. 
        Results are shown for analyses with two different QNM mode contents. 
        The gray region shows the expected exponential decay, calculated from the inferred remnant mass and spin, referenced to the start time indicated by the vertical line.
        The error bars and the gray regions show the 50\% and 90\% uncertainties.
        \emph{Bottom panel:} the amplitude of the ${(2,2,1)}$ overtone. 
        Results are shown for the overtone analysis and the gray region shows the expected decay referenced to an earlier start time appropriate for the fast-decaying overtone.
    }
\end{figure*}

The GW signatures of merging BSs have been extensively studied using NR: in head-on collisions \cite{Palenzuela_2007, Bezares_2017, Helfer_2022, Evstafyeva_2023, Ge_2025}, quasi-circular inspirals \cite{Bezares_2022, Siemonsen_2023a, Siemonsen_2023b, Evstafyeva_2024_BScatalog}, and scattering interactions \cite{Damour_2025}.
These systems exhibit a rich range of behaviors \cite{Liebling_2012_review, Bezares_2024_review}. 
When the BSs are well separated, they behave as point masses, with details of their internal structure effaced from the dynamics \cite{Damour_1987_effacement}. 
Near merger, the dynamics are affected by tidal deformability (which depends strongly on compactness) and by additional interactions mediated by the overlapping scalar fields.
The latter depend on the relative signs of the frequencies (\emph{regular-} or \emph{anti}-BSs) and the difference between the two phases $\delta\psi$.

Reference \cite{Evstafyeva_2024_BScatalog} presented a catalog of NR simulations \cite{BS_repo} of equal-mass, nonspinning BS binaries spanning different values of $A$ (and hence compactness), phase differences, and combinations of regular- and anti-BSs, using multiple NR codes. We use the \texttt{A17-d17} \texttt{\textsc{GRChombo}} \cite{Clough_2015_GRChombo, Radia_2022_GRChombo, Andrade_2021_GRChombo, GRTLCollaboration:2026eza} simulation of two equal-mass \texttt{A17} regular-BSs with zero phase difference.
This zero-phase difference simulation produces only even $m$ GW harmonics, like an equal-mass BH binary, see discussion in Ref.~\cite{Evstafyeva_lessons}.
The BSs start ${d=17}$ (code units) apart on a quasi-circular orbit (eccentricity ${\lesssim 0.005}$) and complete more than 8.5 orbits before merger.
Further details on the initial data and numerical evolution are given in Ref.~\cite{Evstafyeva_2024_BScatalog}.
These BSs are good BH mimickers; the waveform resembles that of a binary BH, see Fig.~\ref{fig:injection}.

During the binary BS inspiral, the objects have no horizons and are surrounded by nonvanishing scalar fields.
However, when these BSs merge, they promptly collapse to form a BH.
At late times, the remnant is expected to settle to a Kerr BH, although scalar-field debris remains near the horizon for several dynamical times after merger.
Thus, the Kerr and vacuum assumptions are strongly violated during the inspiral and early ringdown.

We emphasize that our motivation for studying BSs is not that they are considered likely candidates to describe any current observation. 
Instead, they provide a specific, well-posed example of a non-Kerr-vacuum-GR system amenable to NR simulation.
We use BSs as a case study to investigate how current tests based on Kerr-vacuum-GR binary-BH waveforms respond to a GW signal generated by physics outside these assumptions.

The waveform was scaled to a total binary mass ${M\!=\!71.5M_\odot}$ and resampled to $4096\,\mathrm{Hz}$.
The start of the waveform was tapered, setting the low-frequency cutoff ${f_{\rm low}=23.4,\mathrm{Hz}}$ used in all analyses.
The signal was injected into two detectors (H1, L1) at a luminosity distance of ${425\,\mathrm{Mpc}}$, chosen to yield a signal-to-noise ratio (SNR) of ${\rho_{\rm network}=80.2}$, matching GW250114.
All other parameters (including the peak time, $t_{\rm peak}$) were set to the values in Tab.~2 of the supplement of Ref.~\cite{GW250114_area_test}.
Simulated noise, similar to that around GW250114, was generated from the representative amplitude spectral density \texttt{aLIGO\_O4\_high\_asd}.

\begin{figure*}[t]
    \centering
    \includegraphics[width=0.96\textwidth]{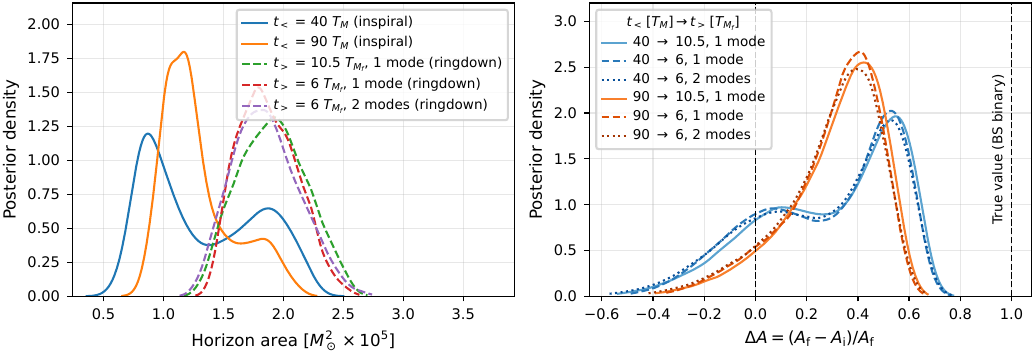}
    \caption{ \label{fig:area_test} 
        The area-law test.
        \emph{Left panel:} The total horizon area (in units of square solar mass) from the two inspiral (solid lines) and the three ringdown (dashed lines) analyses.
        The posteriors for the initial area inferred from the inspiral analyses show multimodality, associated with the ridge-like structures seen in the mass posteriors in Fig.~\ref{fig:inspiral}.
        \emph{Right panel:} The change in area across the merger for the combination of inspiral and ringdown analysis indicated in the legend.
        The true horizon area increase in the BS injection is indicated by the rightmost vertical dashed line.
        This should be compared with Fig.~5 in Ref.~\cite{GW250114_area_test}, except here the change is normalized to the final area because the initial horizon area is zero for the BS inspiral.
        For every analysis combination the area-law test is passed, $\Delta A>0$, with confidences ranging from $80\%$ to $92\%$.
    }
\end{figure*}

It has been shown that the \texttt{A17-d17} simulation can be fit using binary BH waveforms \cite{Evstafyeva_2024_BScatalog}, albeit with parameter biases, despite producing signs of apparent deviations from GR in some standard tests \cite{Evstafyeva_lessons}.
However, those analyses used full inspiral-merger-ringdown models in the frequency domain at lower SNRs of ${20 \lesssim \rho_{\rm network} \lesssim 70}$.
This Letter analyzes the inspiral and ringdown separately in the time domain at higher GW250114-like SNRs.
An analysis of the entire signal is shown in the Supplement.

{\textit{\textbf{Inspiral Analysis}}---}Following Ref.~\cite{GW250114_area_test}, the signal was split into inspiral and ringdown portions, with a gap around merger.
The inspiral ends at time $t_<$; two values were considered, ${t_{\rm peak}-t_<=40 \, T_M}$ and ${90 \, T_M}$, where ${T_{M}=0.354\,\mathrm{ms}}$ is the time unit associated with the total mass of the binary (see Tab.~2 of the supplement of Ref.~\cite{GW250114_area_test}). 
These times correspond approximately to ${\sim 1.5}$ and $2.5$ GW cycles before the peak strain.

The inspiral data were analyzed between $f_{\rm low}$ and ${f_{\rm Nyquist}=1024\,\mathrm{Hz}}$ using the NR surrogate model \texttt{\textsc{NRSur7dq4}} \cite{NRSur7dq4} with \texttt{\textsc{TDInf}} 
\cite{tdinf, GW150914_area_test, 2024PhRvD.109b4024M, 2025PhRvD.112j4046M}.
TDInf uses a time-domain log-likelihood which allows the analysis data to be truncated precisely at ${t_<}$.
The source sky position and merger time were fixed to the injected values while the other parameters were sampled using the priors described in Tab.~\ref{tab:insp_priors}.
\texttt{\textsc{TDInf}} internally applies additional parameter transformations to improve sampling efficiency.

The posterior was sampled using \texttt{\textsc{emcee}} \cite{emcee} with 64 walkers initialized from the prior.
The walkers were evolved for a ${2 \times 10^4}$-step burn-in, after which each walker was scored based on its median log posterior of the preceding ${2 \times 10^3}$ steps. 
The 20 lowest-scoring walkers were then reseeded as perturbed, randomly chosen copies of the remaining walkers. 
The ensemble was evolved for a further ${5.8 \times 10^5}$ steps and the resulting chains were thinned by a factor of $10^4$ chosen to be larger than the measured autocorrelation length. 
This left $3712$ effectively independent posterior samples.
The long chain lengths and aggressive thinning were necessary because of the posterior structure described below.

The results of the inspiral analysis are summarized in Fig.~\ref{fig:inspiral}.
The posterior has a complicated structure, particularly in the component masses.
The two-dimensional component-mass posterior consists of multiple narrow ridges that roughly lie along contours of constant chirp mass.
This unusual posterior geometry was also found in the full-signal analysis described in the Supplement where it is discussed further.
The mass ratio is poorly constrained with the posterior having support all the way to the lower limit of the prior.
The posterior geometries for the remaining parameters are simpler, although we note that those for the component spins are broader than for GW250114 and include support at large values.

{\textit{\textbf{Ringdown Analysis}}---}The ringdown begins at time $t_>$; values of ${t_>-t_{\rm peak}}$ ranging from $1.0$ to ${6.7\,\mathrm{ms}}$ were considered.
The ringdown was analyzed using \texttt{\textsc{ringdown}} \cite{Ringdown}.
This also uses a time-domain implementation of the log-likelihood and can include different combinations of QNMs, whose Kerr frequencies are computed from the remnant mass and spin using \texttt{\textsc{qnm}} \cite{Stein_QNM}.
Following Ref.~\cite{GW250114_area_test}, results are shown for a one-mode analysis using the fundamental $(\ell,|m|,n)=(2,2,0)$ QNM and a two-mode analysis including the $(2,2,1)$ overtone.

The sky position was again fixed to the injected value.
The sampling over the QNM amplitudes $A$ and phases $\phi$, and the remnant mass and spin, was performed using the \texttt{\textsc{numpyro}} \cite{numpyro} No-U-Turn Sampler (NUTS) with four chains evolved for 2000 steps after a 1000-step burn-in and convergence was assessed using the effective sample size and the Gelman-Rubin statistics.
Uniform priors were used for the final mass, spin, and phases over the ranges ${50 \! < \! M_f/M_\odot \! < \! 85}$, ${0.255 \! < \! \chi_f \! < \! 0.95}$, and ${0 \! \leq \! \phi \! < \! 2\pi}$.
The prior on the amplitude quadratures was Gaussian with standard deviation ${A_{\rm scale}=5\times 10^{-20}}$.

The results of the ringdown analysis are shown in Fig.~\ref{fig:ringdown}.
In contrast to the inspiral, the ringdown posteriors are qualitatively similar to those obtained for GW250114.
This is not surprising because the BSs collapse promptly to a BH after merging, making the postmerger substantially closer to Kerr-vacuum-GR.

{\textit{\textbf{Area-Law Test}}---}The parameters inferred by the inspiral analysis were used to calculate the initial total horizon area, ${A_i\!=\!A_1+A_2}$, using Eq.~\ref{eq:Kerr_horizon_area}.
Despite no horizons existing during the inspiral, the fitted model infers a nonzero $A_i$; see the left panel of Fig.~\ref{fig:area_test}.
Similarly, the final horizon area, $A_f$, was calculated from the remnant parameters inferred by the ringdown analysis.
This was done at the two ringdown start times indicated in Fig.~\ref{fig:ringdown}.

The results of the area-law test are shown in the right panel of Fig.~\ref{fig:area_test}.
The inferred horizon area increases across the merger, with the dimensionless ${\Delta A=(A_f-A_i)/A_f}$ measured to be positive with a high confidence of $80\%$ to $92\%$ when using the later and earlier $t_<$ respectively.

{\textit{\textbf{Discussion}}---}Area-law tests infer horizon areas using models that assume Kerr BHs in vacuum GR.
This makes their interpretation subtle and model dependent.

The horizon areas are not measured directly from the signal; they are instead derived from object masses and spins inferred under these assumptions.
This does not invalidate the test, but it limits what can be concluded from passing it.
As we have shown, it is possible for other compact objects besides BHs to pass area-law tests.
Passing such tests should be interpreted as a consistency check of the joint Kerr-vacuum-GR assumptions, rather than as direct evidence for any horizon properties.

This Letter has shown that merging BSs can mimic merging BHs and pass an area-law test at high confidence.
We note, however, that the confidence is not as high as that found for the GW250114 analysis, where $\Delta A$ was found to be positive with ${\sim \! 4.4\sigma}$ confidence \cite{GW250114_area_test}. 
Despite passing the area-law test, the BS signal still exhibits anomalies that show up in other analyses that distinguish it from a standard binary BH \cite{Evstafyeva_lessons}. 
This highlights the importance of designing complementary tests and not being overly reliant on any single test of GR.

{\textit{\textbf{Data Availability}}---}The data and code to reproduce all of the analyses here are available on GitHub \href{https://github.com/selmavangstein/horizon-area-law-bosonstar}{\faGithub}.

{\textit{\textbf{Acknowledgments}}---}The authors thank R.\ Dyer, T.\ Evstafyeva, T.\ Jain, T.\ Kinowski, G.\ A.\ Marks, N.\ Sarin, U.\ Sperhake, S.\ Staelens for useful discussions.
\texttt{\textsc{bilby}} \cite{bilby_paper, bilby_doi} was used for the injections.
SV acknowledges support from the Aker Scholarship and the UKRI Frontier Research Guarantee [EP/X035344/1].

\bibliographystyle{apsrev4-1}
\bibliography{references}

\setcounter{equation}{0}
\setcounter{figure}{0}
\setcounter{table}{0}
\makeatletter
\renewcommand{\theequation}{S\arabic{equation}}
\renewcommand{\thefigure}{S\arabic{figure}}
\renewcommand{\thetable}{S\arabic{table}}

\onecolumngrid
\newpage
\begin{center}
  \textbf{\large{Supplemental Material}} \\
\end{center}
\twocolumngrid

{\textit{\textbf{Inspiral Analysis Priors}}}---The parameter prior distributions used for \texttt{\textsc{TDInf}} analysis are given in Tab.~\ref{tab:insp_priors}. 

\begin{table}[h]
    \caption{\label{tab:insp_priors}
        Priors used for the \texttt{\textsc{TDInf}} inspiral analysis are of the same functional form as those used in Ref.~\cite{GW250114_area_test}.
    }
    \begin{ruledtabular}
        \begin{tabular}{lcr}
            Parameter & Range & Prior \\
            \hline 
            Total binary mass & $M\in[58,82]\,M_\odot$ & Uniform \\
            Binary mass ratio & $q\in[0.25,1]\,$ & Uniform \\
            Luminosity distance & $d_{\rm L}\in[200,700]\,\mathrm{Mpc}$ & Uniform \\
            Spin magnitudes & $|\vec{\chi}_1|, |\vec{\chi}_2|\in[0, 0.9]\,$ & Uniform \\
            Spin tilts & $\theta_1, \theta_2\in[0, \pi]\,$ & Uniform in cosine \\
            Spin separation & $\phi_{12}\in[0, 2\pi)\,$ & Uniform \\
            Spin azimuth & $\phi_{\rm JL}\in[0, 2\pi)\,$ & Uniform \\
            Phase & $\phi\in[0,2\pi)$ & Uniform \\
            Zenith angle & $\theta_{\rm JN}\in[0,\pi]$ & Uniform in cosine
        \end{tabular}
    \end{ruledtabular}
\end{table}

\begin{figure}[h]
    \centering
    \includegraphics[width=\columnwidth]{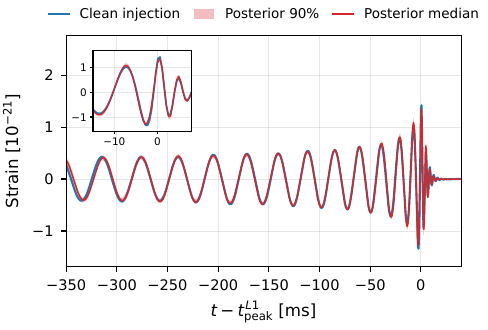}
    \caption{ \label{fig:full_signal_reconstruction}
        Full signal reconstruction. 
        The injected BS signal in the L1 interferometer is shown along with the 90\% credible region on the reconstructed signal from the full inspiral-merger-ringdown analysis.
    }
\end{figure}

\begin{figure}[b]
    \centering
    \includegraphics[width=\columnwidth]{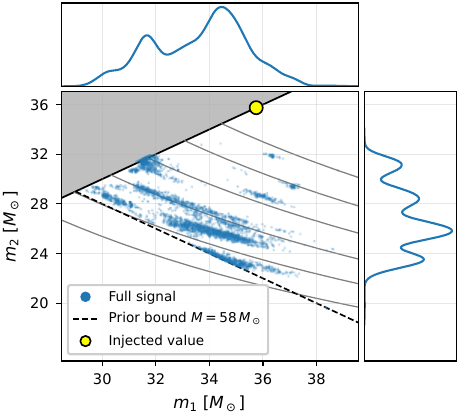}
    \caption{ \label{fig:full_signal}
        The full signal analysis.
        Posteriors from the full inspiral-merger-ringdown analysis on the component masses are shown.
        These should be compared with those from the inspiral analysis in Fig.~\ref{fig:inspiral}.
        Analyzing the full signal leads to a more informative measurement of the mass ratio.
        Some signs of the ridged structure in the posterior on the component masses are still present.
    }
\end{figure}

{\textit{\textbf{Inspiral-Merger-Ringdown Analysis}}---}This section contains the results of an inspiral-merger-ringdown analysis of the full BS injected signal.
For simplicity, and to be consistent with the inspiral and ringdown analyses described in the main text, the sky position was fixed to the injected value.
The analysis was performed using \texttt{\textsc{TDInf}} with the same settings as described in the main Letter, but with the end time pushed past the duration of the signal.
The priors were also the same as used for the inspiral analysis (see Tab.~\ref{tab:insp_priors}).
The waveform reconstruction is shown in Fig.~\ref{fig:full_signal_reconstruction} and posteriors on the component masses are shown in Fig.~\ref{fig:full_signal}.

The posterior geometry on the component masses has a much more complicated structure than is usually found in GW inference for binary BH signals.
We conjecture that this is due to a combination of several unusual features of this analysis.
Firstly, the injection has a very high SNR, chosen to match GW250114.
Secondly, the sky position and merger time are fixed during this \texttt{\textsc{TDInf}} analysis.
Thirdly, and this is likely the most important feature, there is (by design) severe model misspecification present in this analysis because of the way a binary BS injection it fitted using a binary BH model; this is further compounded by the previously noted high SNR.

\newpage

\end{document}